# Metastable Multi-centered Polarons in BiVO$_4$

*Seyeon Park$^{a,‡}$, Yajing Zhang$^{a,b,†,‡}$, Michele Reticcioli$^{c,d,e}$, Cesare Franchini$^{c,f}$, and Bongjae Kim$^{a}$\**

$^a$Department of Physics, Kyungpook National University, Daegu 41566, Korea

$^b$Max Planck Korea/POSTECH Center for Complex Phase Materials, Pohang 37673, Korea

$^c$Faculty of Physics and Center for Computational Materials Science, University of Vienna, Vienna 1090, Austria

$^d$National Research Council, CNR-SPIN, L'Aquila 67100, Italy

$^e$University of L'Aquila, L'Aquila 67100, Italy

$^f$Department of Physics and Astronomy "Augusto Righi," University of Bologna, Bologna 40126, Italy

\*Email: bongjae@knu.ac.kr

KEYWORDS: Polaron, BiVO$_4$, DFT, optical experiment

ABSTRACT: Polarons, quasiparticles formed through interactions between lattice and charge carriers (electrons and holes), strongly influence the electronic and optical properties of functional materials. In nanostructured $BiVO_4$, polaron formation and dynamics govern photocatalytic efficiency and charge transport, yet the microscopic nature remains not fully resolved. Here, using first-principles calculations, we report the formation of multi-centered polarons, in contrast to the more common single-centered states. Moreover, electron polarons exhibit pronounced anisotropy compared to the isotropic hole counterpart, reflecting a distinct character in charge-lattice coupling. These theoretical insights offer a direct interpretation of optical and spectroscopic experiments, providing strong evidence of anisotropic multi-centered polaronic behavior in $BiVO_4$. The presence of multiple in-gap states, especially from multi-centered polarons, introduces new channels for charge transport and recombination, possibly offering opportunities to control carrier dynamics in nanoscale photocatalytic and optoelectronic devices.

1. INTRODUCTION

In polarizable solids, excess charge carriers can couple to local lattice distortion and form self-trapping quasiparticles known as polarons. These spatially localized states, often induced from the short-range electron-phonon interactions, are called Holstein small polarons, and they significantly alter the electronic properties of the host materials, including charge mobility, transport properties, catalytic activity, and photochemical effects.[1-11] The formation of such polarons is particularly important for nano-structured materials used in light-energy applications, as they can fundamentally reshape the structure of the optical gap by the formation of in-gap states, thus influencing light absorption.[3, 12, 13] Furthermore, photogenerated charges can interact with localized polarons, limiting the charge transport efficiency which becomes especially pronounced at the nano-scale interfaces.[9, 14-16] Therefore, the correct characterization of the polarons is of fundamental importance for advancing nano-enabled optoelectronic and photocatalytic devices.

Various techniques can be employed to identify the signatures of polarons. As polarons typically form sparsely and locally, detecting them requires sensitive approaches. From the experimental side, scanning tunneling microscopy/spectroscopy (STM/STS) and atomic force microscopy (AFM) are representative techniques that can directly measure lattice deformation and associated local charge responses, which are usually manifested at the surface in nanostructured systems.[17-21] For the polarons in bulk systems, optical measurement and spectroscopies are particularly effective, where they can resolve the narrow energy states arising from the localized polaronic states.[22-31]

From a theoretical side, density functional theory (DFT) has provided essential insights into polaron physics, shedding light on experimental observations. Being a material-specific theory, DFT offers a microscopic explanation for the target system and its material properties. In particular, for the photocatalytic and photochemical activities, the combination of DFT and experimental tools has proven highly successful in elucidating the fundamental role of the polarons in photochemical processes.[3, 8, 12, 13, 17, 19, 32-35] This microscopic understanding is crucial to advancing the efficiency of devices in technological applications such as water splitting.[9, 36, 37]

Traditionally, the theoretical modeling of polarons is based on the idea that a single excess charge carrier couples with a local lattice deformation, forming a single quasi-particle state localized on a single hosting site. However, recent reports suggest that fractionalized excitations - interpreted as fractional charges - can emerge[38-40] and may hold a potential for future quantum computing applications.[41] Polarons can exhibit similar fractionalization, where a single excess charge carrier can simultaneously couple to multiple local lattice deformations, in the form of the weakly localized identity, known as multi-centered polarons.[13, 21, 42-46] These fractionally charged states are distinct from the multi-polaron states, which involve interactions among whole-charged, single-site polarons.[47-50] With recent advances in experimental techniques enabling not only the detection but also the direct creation and manipulation of polarons at the nanoscale,[21] the concept of fractionalized polarons offers new dimensions of control and functionality.

In this study, we report the formation of multi-centered polarons in $BiVO_4$. The material has been the subject of a wide range of theoretical and experimental investigations on polaron properties,[3-5, 24, 32-35, 51-54] with even more extensive studies on practical applications such as photocatalysis and water-splitting.

Employing a hybrid functional DFT approach, our computations demonstrate that multi-centered polarons can arise from a single excess charge in $BiVO_4$. Our results show that these multi-centered electron polarons form in-gap states just below the conduction states, occurring at much higher energies than more common single-center polaron states. The accuracy of our predictions is reflected in the excellent agreement between our calculated dielectric function and experimental optical measurements. Importantly, we show that the electronic anisotropy of the electron polaron is preserved in multi-centered polaron configurations, which is consistent with the spectroscopic studies. Finally, we describe the fundamental properties of the hole polaron counterpart.

## 2. RESULTS AND DISCUSSION

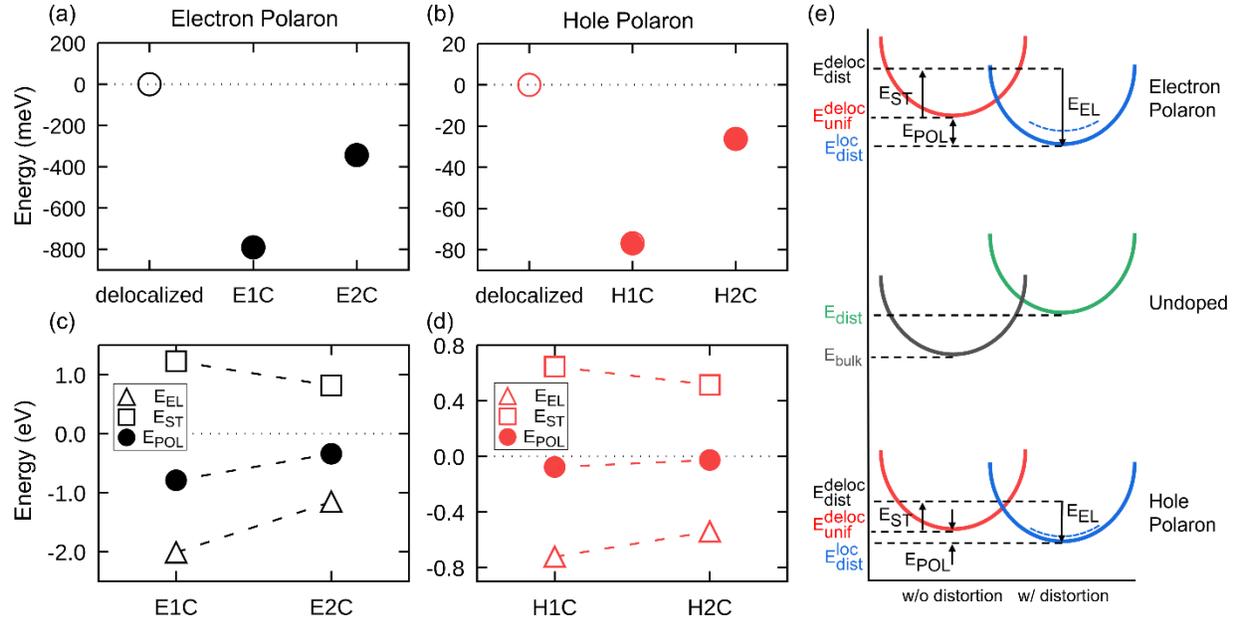

**Figure 1.** Energetics of delocalized, one-center (E1C and H1C), and two-center (E2C and H2C) polaron states in BiVO$_4$ for excess (a) electron and (b) hole. (c) and (d) denotes corresponding polaron formation energy ($E_{POL}$), electronic energy gain ($E_{EL}$), and structural energy loss ($E_{ST}$) for electron and hole polaron, respectively. (e) shows a schematic energy diagram of the polaron formation process.

Once an excess charge is inserted into an insulating system like BiVO$_4$, the charge can be either delocalized to generate a metallic phase or form a small polaronic insulating phase. In DFT-based approaches, the stability of the polaronic phase can be determined by directly comparing the total energy of polaronic and delocalized solutions. In BiVO$_4$, both electron and hole polaronic solutions are found to be lower in energy than the delocalized cases,[32, 53, 55] with the excess charge carrier, either an electron or a hole, reported to localize at specific, single atoms (the V-site for electron and the Bi-site for hole polaron).

Interestingly, we found that different polaronic solutions can be stabilized in the DFT calculation, including one-centered and two-centered polarons. In Figure 1a,b, we show the relative energies of the electron/hole one-centered (E1C/H1C) and two-centered (E2C/H2C) polarons. First, one can see that single-center polaronic solutions, E1C and H1C, are stable as found in previous studies.[3, 32, 33, 37, 56-58] As compared to the two-centered states, they are lower in energy for both the electron and hole cases. The E1C polaron is approximately 0.8 eV lower in energy than the delocalized case, whereas the H1C polaron is only 80 meV more stable than the metallic *p*-doped solution. This difference of one order of magnitude in polaron formation energy seems to arise primarily from the electronic energy gain $E_{EL}$: electrons localizing on V-*d* orbitals gain 2 eV as compared to the delocalized solution, while hole localization on Bi-*s*/*p* orbitals gives rise to a gain of only 0.7 eV.

Other than these conventional one-site-centered polarons, we can also stabilize the multi-centered polarons. As shown in Figure 1a,b, the two-centered polarons, E2C and H2C, are not the ground state solution of the system; yet, they represent a metastable phase that is lower in energy than the delocalized solution. As we will discuss later, comparisons with experimental data strongly advocate the formation of multi-centered polarons. We were able to stabilize different metastable solutions with similar electronic characteristics, which are very close in energy, and we envisage a variety of such states. The detailed analysis of different multi-centered polaron configurations is presented in the Supporting Information with Figure S2. We note that the calculated stability of the multi-centered polarons depends on the computational setup, which is analyzed in detail in Supporting Information with Figure S1.

For the electron polarons, we found the local moment for E1C is 0.96 $\mu_B$, which corresponds to a single spin from one excess charge at the vanadium site. In contrast, for E2C, we identified a spin moment of 0.46 $\mu_B$ at V-site, which indicates the single electronic charge is shared between two vanadium centers in the form of quasiparticles. This indeed demonstrates the split of the single electron charge into two V-sites. Here, we note that the hosting atoms of the partial charge are not necessarily close to each other in space. Various combinations of two V pairs can equally host multi-centered polarons, and are very close in their energies (See Figure S2a in Supporting Information).

Hole polarons can also host fractional charge polarons, but, due to their extended character, the spin moments are distributed across oxygen sites surrounding Bi atoms. We found spin moments of 0.098 $\mu_B$ and 0.042 $\mu_B$ for H1C and H2C at Bi-site, respectively. Here, nearby oxygen atoms carry approximately 0.041 $\mu_B$ and 0.076 $\mu_B$, which are much larger than the fractional charges of 0.014 $\mu_B$ and 0.0075 $\mu_B$ for oxygen near V atom sites for E1C and E2C polarons. We note again that this polaronic phase with fractional charge is different from the fractional charges obtained from excited quasiparticles, and is static in nature, enabling direct manipulation within the nanoscale tools.[21, 42]

To understand the microscopic formation mechanism of multi-centered polarons, we analyzed the polaron binding energy ($E_{POL}$) in detail, as shown in Figure 1c–e.[59] As schematically depicted in Figure 1e, $E_{POL}$ is determined by the electronic energy gain, $E_{EL}$, from the charge localization, which is partially compensated by the energy loss due to the structural deformation, $E_{ST}$. For the undoped case, a distorted structure is always higher in energy and cannot be stabilized. Once a charge is doped, a polaronic state is stabilized when the $E_{EL}$ is larger than the $E_{ST}$, as depicted in Figure 1e.

As electron polarons are more localized due to V-$d$ character, $E_{EL}$ is much larger with stronger structural distortion; hence, $E_{ST}$ is also larger than in the hole polaron cases. For the E1C case, we found that the structural energy loss is $E_{ST}$ = 1.22 eV and the electronic energy gain is $E_{EL}$ = 2.01 eV. This overall energetics is similar for multi-centered polarons. In E2C, we notice the charge is split and distributed to two V-centers, hence the degree of charge localization and associated structural distortions are softened compared to E1C (See Figures S3 and S4). Thus, as shown in Figure 1a,c, $E_{ST}$ and $E_{EL}$ are decreased, and $E_{POL}$ is smaller than E2C. Noteworthy is that as the charge is distributed into two sites, reduced localization affects the electronic energy gain, $E_{EL}$, more than the structural distortion, $E_{ST}$; hence, the E1C is more favorable than the E2C.

Similarly, except for the smaller energy scales, we observe that the exact mechanism is in action for hole polarons. For H1C, due to the less localized $s/p$ character of the polaron, the energy gain from the charge localization is only a few tens of meV higher with $E_{ST}$ = 0.65 eV and $E_{EL}$ = 0.72 eV, and $E_{POL}$ is 0.08 eV, which is one order smaller than the E1C cases.[32, 55] The energy difference between $E_{EL}$ and $E_{ST}$ becomes even smaller for H2C, and we expect that, compared to E1C, multi-centered hole polaronic states with a higher number of centers are not readily formed, compared to the electron cases (See Figures S3 and S4)

Due to the existence of multi-centered polarons, as schematically shown in Figure 1e, different energy states among polaronic states are possible (see dashed blue curves). Such localized states directly affect charge excitations from the valence to the conduction bands, as they form in-gap states with various energy levels. As the temperature can relocate the energetic positions of polaronic states, we expect that diverse channels of charge transport exist at the interfaces.[3] Optical measurement is one of the ideal tools to capture multi-centered polaronic states, which will be discussed in more detail later.

Note that in previous hybrid functional studies on polarons, the employment of the appropriate fraction of HF exchange has been important for the better description of the electronic structures.[32, 55, 58, 60-64] This also applies to the cases of the multi-centered polarons such that the amount of fraction of exact exchange, α, in the HF exchange is found to be essential for the stability of the various multi-centered polarons phases (Figure S1 in Supporting Information), and this plays a vital role in the determination of the charge localization and accompanying distortions, as seen in Figures S3 and S4. However, we also note that the qualitative results we have discussed remain solid for this HF mixing. We expect other advanced computational approaches, such as GW and DFT+$U$+$V$, can be further employed in the future.[65, 66]

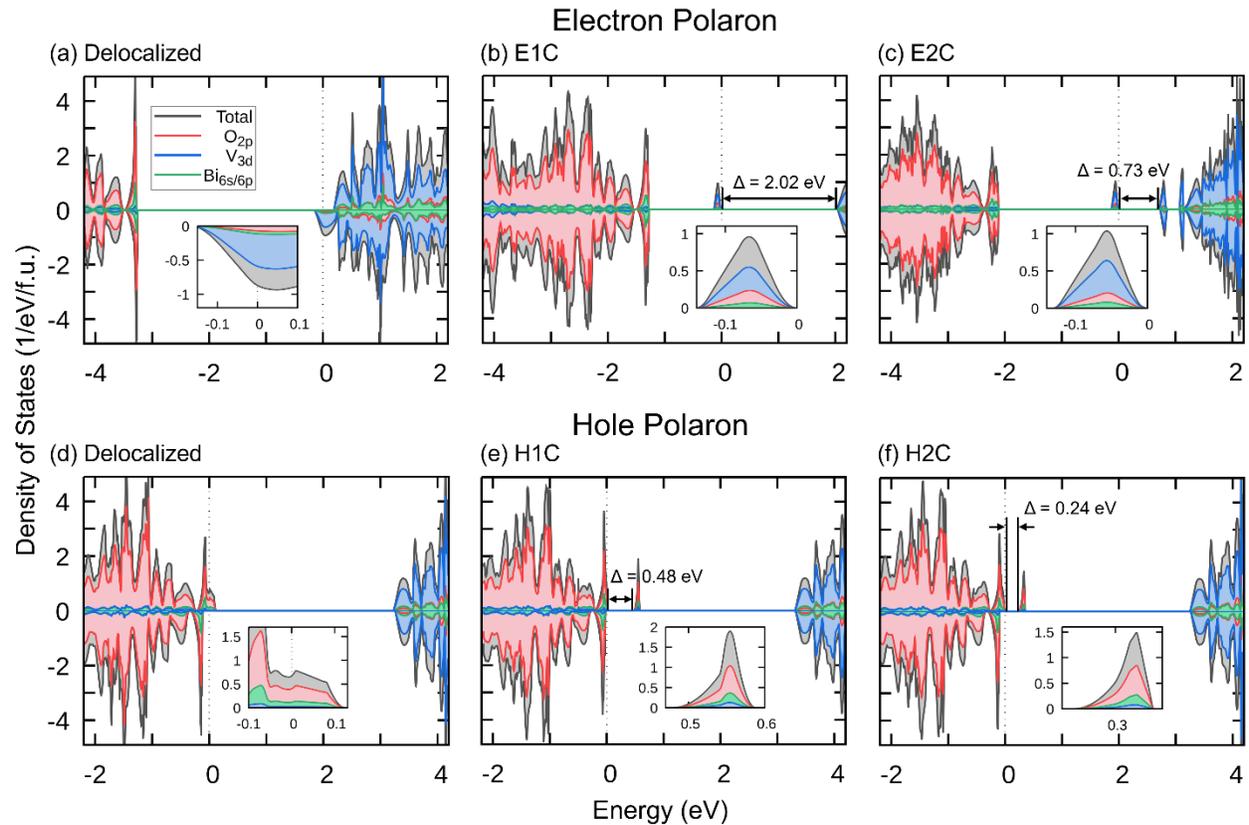

**Figure 2.** Density of states (DOS) of the electron- and hole-doped BiVO$_4$. Black, red, blue, and green lines represent the total DOS, O 2*p*, V 3*d*, and Bi 6*s*/6*p* orbitals, respectively. (a–c) correspond to electron doping (delocalized, E1C and E2C), and (d–f) to hole doping (delocalized, H1C and H2C). The energy separation between the conduction band and mid-gap polaron states is denoted as Δ.

A stark contrast in the electronic structure is evident in the polaronic in-gap states between the multi-centered and single-centered configurations. In Figure 2, we display the density of states (DOS) for the delocalized, single-centered, and two-centered polarons in electron and hole cases. The partial DOSs for delocalized cases (Figure 2a,d) clearly show metallic behavior, with the excess electrons and holes rigidly shifting the Fermi energy. For pristine BiVO$_4$, the calculated energy gap is 3.19 eV, which remains essentially unchanged upon simple charge doping, except for the overall downward shift of the overall electronic states. Once polarons are formed, however, the system remains insulating, with the emergence of localized in-gap states (See Figure 2b,c).

Significantly, the energy position of the multi-centered polaronic states differs markedly from that of the single-centered ones. In the electron polaron case, E1C exhibits an in-gap state which is well separated from both conduction and valence bands (Figure 2b). In contrast, for E2C (Figure 2c), the localized states lie much closer to the conduction band with an energy separation Δ of about 0.73 eV. These E2C polaronic states are higher in energy than the E1C states, forming a metastable phase as indicated by the energetics (Figure 1a). We expect that the different energy positions of the diverse polaronic states can be directly probed via experimental techniques such as optical measurements.[67, 68] The expected V-*d* character of the polarons for both E1C and E2C is illustrated in the insets of Figure 2b,c.

H1C and H2C also have distinct energy positions for the hole polaron cases. In the delocalized cases, the excess hole induces a simple rigid shift as shown in Figure 2d, while in the polaronic case, it forms localized in-gap states. These states are much closer to the valence bands, and, combined with the extended $s/p$ characteristics, the polaronic solutions are energetically closer to the delocalized case, compared to the electron polaron counterparts (see Figure 1a,b). Unlike the electron polaron cases, we see that the multi-centered polaron states are closer to the valence band than the single-charge one, and we expect such states to merge with the delocalized solutions eventually. Due to the extended character of the hole polarons (Figure S5d–e), hybrid functional calculations, where exact exchange and both short-range and long-range interactions are treated, are essential here, and methods such as DFT+$U$ may result in misleading phases.[58, 61]

The positions of the in-gap polarons are crucial for efficiently determining photocatalytic activity.[3, 12, 34, 69, 70] In BiVO$_4$, the charge transport, primarily governed by hopping dynamics, is intrinsically set by the energy levels of the localized polarons. Moreover, polarons can hinder electron transport on the macroscale and are thus highly detrimental to the performance of nanostructured devices.[35] In water-splitting applications, the energy levels of the electron and hole polarons govern the recombination process, which competes with charge transfer to the electrolyte during redox reactions.[3] Our findings of the multi-centered electron (hole) polarons, which lie at higher (lower) energies within the gap, imply that alongside the stable single-centered polaronic states, multiple channels may exist for the charge transport and recombination. This finding can offer a new strategy to enhance the efficiency of BiVO$_4$-based photocatalytic devices by activating specific polaronic channels at interfaces and devices.[71] Here, we note a recent study on the quasi large-polaron phase in BiVO$_4$, which suggests even more energy levels.[72]

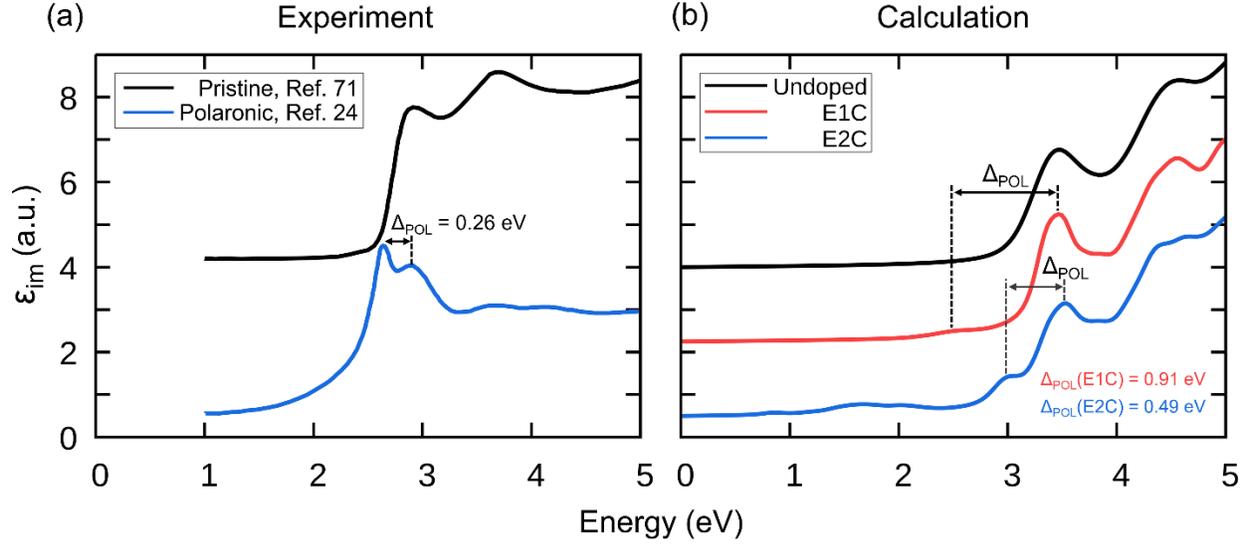

**Figure 3.** (a) Experimentally measured imaginary part of the dielectric function $\varepsilon_{im}$ of BiVO$_4$ in pristine and polaronic phases, adapted from previous studies.[24, 73] (b) Calculated $\varepsilon_{im}$, for pristine (black), E1C (red), and E2C (blue) polaron cases. $\Delta_{POL}$ indicate the energy separation between the main and sub-polaronic peaks in experiments and calculations.

Optical measurements provide clear evidence for the presence of multi-centered polarons. In Figure 3, we compare the experimental and calculated dielectric functions from polaronic and non-polaronic phases of BiVO$_4$.[24, 73] In Figure 3a, we replot the experimental imaginary part of the dielectric function, $\varepsilon_{im}$, for the monoclinic and orthorhombic BiVO$_4$ corresponding to pristine and polaronic phases, respectively, as reported previously.[24, 28] The main transitions from the valence to conduction bands yield a pronounced optical peak at 2.91 eV in the non-polaronic phase. A distinct double-peak structure is exhibited in the polaronic phase, with a sharp additional sub-peak at 2.65 eV just below the main optical feature. The position of this polaron-induced peak highlights key essential characteristics of the polaron states.

Our optical calculations unambiguously identified the origin of the experimental optical features. For the pristine case, our calculated $\varepsilon_{im}$ accurately reproduces the optical gap and the corresponding central peak, albeit it slightly overestimates the peak position at 3.46 eV (Figure 3b), about 20% larger than the calculated fundamental gap.

For the E1C case, our $\varepsilon_{im}$ differs from that of the pristine case. The mid-gap single-charge polaronic state (Figure 2b) induces a broad additional hump at around 0.91 eV (see $\Delta_{POL}$(E1C) in Figure 3b) below the central peak, while preserving the overall curve shape. However, this broad bump from E1C fails to account for the sharp additional peak from the experiment.

In the E2C case, by contrast, our dielectric function calculation correctly reproduces the experimentally observed double-peak structure. Here, the polaronic states are closer to the conduction bands (Figure 2c), leading to an optical feature that closely matches the experimental observations. The separation between the two peaks, $\Delta_{POL}$(E2C) is about 0.49 eV, which is comparable to the corresponding $\Delta$ of E2C value (0.73 eV in Figure 2c), and is much closer to the experimental value of 0.26 eV.

We note that for the hole-polaron cases, whether they are single- or multi-polarons, the polaronic in-gap states are formed just above the valence bands, and the corresponding optical response is reflected as a lower energy peak, as shown in Figure S6, meaning the experimental double-peak feature in Figure 3a is not from the hole-polarons.

While the gap size and the polaronic peak positions are sensitive to HF mixing ratio (Figures S6), and could be further refined by parameter tuning or more advanced computation methods, our study essentially provides strong evidence for forming a multi-center polaron.[65] This finding contrasts with previous studies in which the single-charge electron polaronic state lies deep inside the gap,[3, 32] and the hole polaronic state is positioned just above the valence band maximum,[55, 63] neither of which accounts for the experimental two-peak structure.

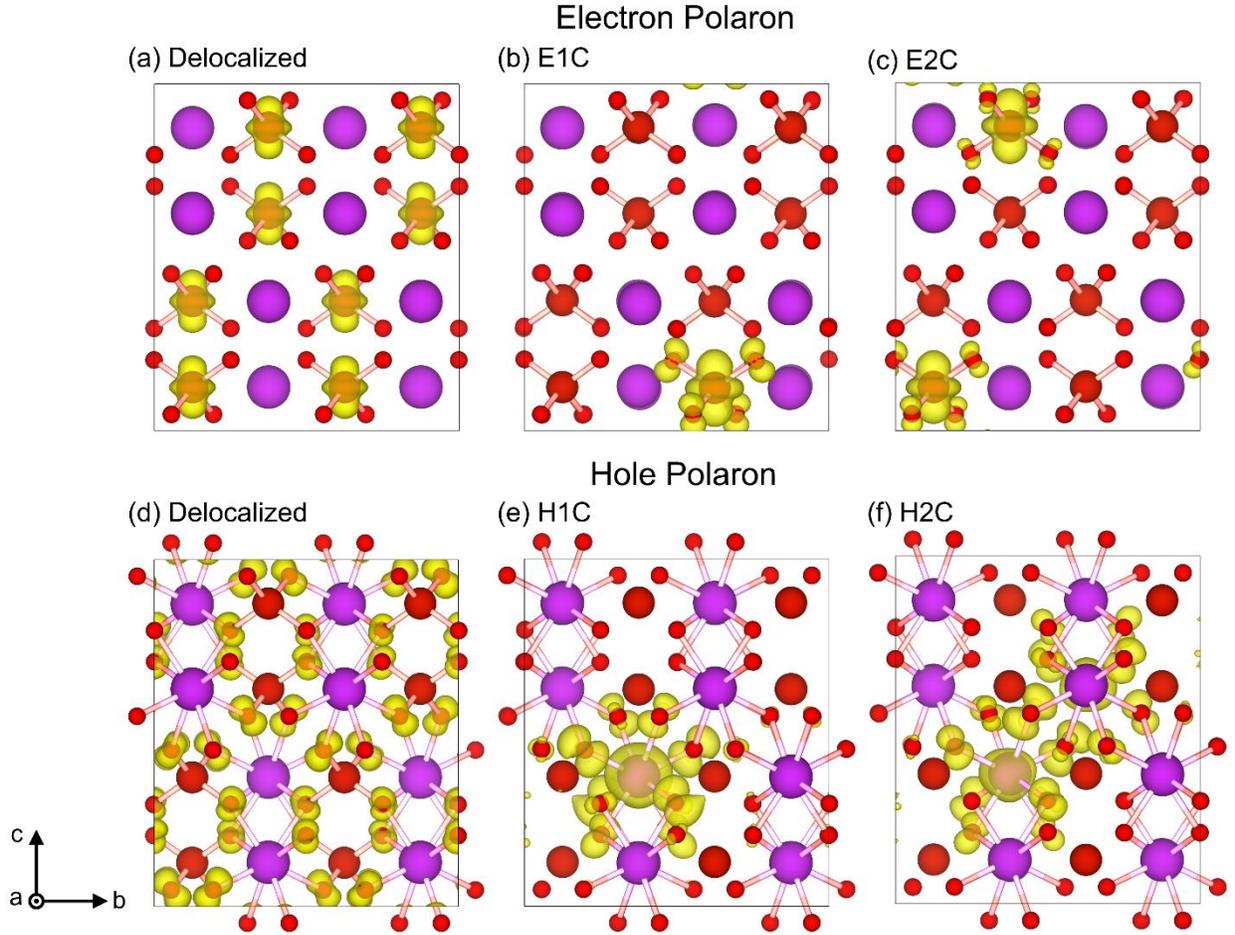

**Figure 4.** Partial charge densities (PCDs) of BiVO$_4$, (a–c) electron and (d–f) hole doping. (a) and (d) are delocalized electron and hole, respectively. (b,e) and (c,f) are one (E1C and H1C) and two (E2C and H2C) centered polaron, respectively

Figure 4 presents the partial charge densities (PCDs) for both electron- and hole-doped cases. The excess charge primarily localizes around the V site upon electron doping, as identified from the DOS in Figure 2. Even in the delocalized metallic case, the excess charges show strong anisotropy, clearly indicating a predominant $d_{z^2}$ character.

For the E1C case, the excess charge is localized at the V site, as shown in Figure 4b, and shared with the four surrounding oxygen atoms within the tetrahedral unit (See also the inset in Figure 2b). The polaron charge distribution at the V site indicates strong anisotropy, a feature that persists in the multi-centered polaron configurations. In the E2C case, although the excess charge is shared between two $VO_4$ tetrahedral units, the PCD retains similar characteristics. The excess charge and correlated distortions are softened for the multi-polaron cases, as shown in Figures S3 and S4.

X-ray absorption spectroscopy has confirmed this anisotropic polaronic nature,[24, 28] where the $d_{z^2}$ orbital-induced anisotropy features are observed in the V-$L_3$ spectrum near the conduction band edge. This feature is critically observed upon the polarization direction of the incident light. As with the dielectric function results, both the energetic position of the polaronic states which is just below the central V-$d$ peak and their strong anisotropy provide categorical evidence for the presence of multi-centered electron polarons in the system. Notably, this anisotropy is not configuration-dependent, as shown in Figure S2d.

In contrast to the electron polaron with the local anisotropic V-$d$ character, the PCD for the hole polaron shows a highly isotropic feature with mixed Bi-$s/p$ and O-$p$ character. This isotropy is also preserved in the various multi-centered hole polaron configurations (See Figure 4d–f).

Here, we briefly discuss the idea of the hole polaron scenario in the explanation of the double-peaked optical feature, as in previous dielectric function and spectroscopy studies.[24, 28] As demonstrated in Figure 2e,f, our calculations show that the localized states arising from the hole polarons are positioned very close to the valence band edge, distinctly different from the electron polaron cases. This means that the resulting optical transitions involving hole polarons produce low-energy features, as seen in Figure 3b (See grey lines for H1C). Given this, we affirm that the polaronic states reported previously,[24, 28] are more consistently attributed to the electron polaron. Specifically, the multi-centered electron polarons described above, rather than to hole polarons are responsible. This is further supported by the clear anisotropic features reported in the spectroscopic experiment just below the main *L*-edge of the V-*d* states, which align well with our model of anisotropic charge distribution in electron polarons and contrast with the highly isotropic nature expected for hole polarons (Figure 4).

## 3. CONCLUSIONS

In summary, we report the formation of multi-centered polarons in $BiVO_4$, a prototypical material for photocatalysis. Using hybrid-functional DFT, we demonstrated that a single excess charge can be distributed across multiple lattice sites, creating an energetically metastable, but robust multi-centered polaronic phase. These polarons, producing distinct in-gap states that differ markedly from conventional single-centered polarons, form various local in-gap energy states. By directly comparing calculated dielectric functions and partial charge distributions with previously reported optical and spectroscopic measurements, we demonstrate that the existence of multi-centered electronic polarons with strong anisotropy is key to interpreting the experimental observations that had previously been interpreted with a hole polaron picture.

The energetic positions of various in-gap states set the pathways for the charge transport, recombination, and interfacial transfer. The discovery of multi-centered polarons introduces new channels for controlling electron-hole dynamics in nanoscale photocatalytic and optoelectronic devices. Furthermore, as the idea of multi-centered polarons is general and can be applicable to the general complex oxides, tailoring the stability and configurations of the various polaronic states can provide new routes for enhancing the efficiency of not only $BiVO_4$-based devices but also of related oxide measurements.

Finally, we note recent advances on multi-polarons, where the multiple single-charge polarons are stabilized in close proximity despite Coulomb repulsion,[74] demonstrating that the polarons can be added up.[75] In contrast, our present work shows that a single excess charge can be fractionalized into a multi-centered configuration, distributed over different lattice sites. Taken together, these two complementary perspectives, polaron addition and diversion, constitute complementary routes for tailoring local electronic textures in oxides. While the broader implications warrant further experimental tests, these mechanisms provide a practical playground for exploring static fractional charges in solids and directions for engineering novel charge states in quantum technology applications.[38-40]

4. METHODS

**DFT Calculations.** A hybrid functional based on density functional theory (DFT) was adopted to study the polaron properties of $BiVO_4$. A supercell 2×2×1 monoclinic $BiVO_4$ (96 atoms) was created based on the bulk BiVO4. Spin-polarized calculations of supercell $BiVO_4$ were performed using the Vienna ab initio simulation package[76-79] (VASP 6.0.8), employing the Perdew-Burke-Ernzerhof (PBE) exchange combined with exact HF exchange in the HSE06 screened hybrid functionals,[80-82] where only the short-range (SR) exchange is mixed, while the long-range (LR) exchange remains fully GGA.

Both non-spin-polarized and spin-polarized methods were considered for the calculations of BiVO$_4$ systems. Projector augmented-wave (PAW) potentials[83] with valence electrons 6s$^2$6p$^3$ for Bi, 3d$^3$4s$^2$ for V, and 2s$^2$2p$^4$ for O were employed. A cutoff energy of 400 eV was used for the plane-wave basis set. Energy convergence criteria of 10$^{-5}$ eV were set for geometry optimizations and self-consistent energies. A gamma-centered (2×2×2) k-point mesh was used to optimize models and calculate electronic structures of supercells with one electron removed or added.

**Charge carrier doping.** An extra electron is added (electron doping) or removed (hole doping) from the pristine structure to form polarons. A supercell consisting of 16 formula units (f.u.) of BiVO$_4$ was investigated, where the total number of electrons was adjusted from n to n – 1 or n + 1 (from 544 to 543 or 545) to simulate hole and electron doping, respectively. The corresponding charge transitions and polaron site configurations are shown in Supporting Information, Figure S3.

**Lattice distortion.** Polarons induce localized charge around specific central atoms, leading to local lattice distortions: V-O bonds are elongated for electron polarons, while Bi-O bonds are shortened for hole polarons. For electron polarons, V-O bonds around the selected V site were initially stretched before performing full structural relaxation after adding excess charge. For hole polarons, Bi-O bonds were initially shortened around a selected Bi site prior to relaxation. The resulting bond length variations at the polaron sites after relaxation are shown in Supporting Information, Figure S4.

**Energies.** Polaron formation energy E$_{POL}$ is calculated as

$$E_{POL} = E_{dist}^{loc} - E_{unif}^{deloc}$$

where $E_{POL}$ represent the polaron formation energy, $E_{dist}^{loc}$ is the total energy of polaron structure, $E_{unif}^{deloc}$ is the total energy of a delocalized structure including one additional/fewer electron.

$$E_{POL} = E_{EL} + E_{ST}$$

$E_{EL}$ is the electronic energy gain obtained via electronic localization. $E_{ST}$ is the structural energy cost arising from the atomic structure distortions required to accommodate an excess charge (electron or hole).

$$E_{EL} = E_{dist}^{deloc} - E_{dist}^{loc}$$

$$E_{ST} = E_{unif}^{deloc} - E_{dist}^{deloc}$$

$E_{dist}^{deloc}$ is the total energy of the delocalized structure, however, constrained into the atomic structure hosting polaron.[59]

**Dielectric function.** The dielectric function of the systems is directly obtained from DFT results. The imaginary part of the dielectric function was derived from the dynamical response with the Green-Kubo formula:

$$\epsilon_{\alpha\beta}^{(2)}(\omega) = \frac{4\pi^2 e^2}{\Omega} \lim_{q \to 0} \frac{1}{q^2} \sum_{c,v,\mathbf{k}} 2\omega_{\mathbf{k}} \delta(\epsilon_{c\mathbf{k}} - \epsilon_{v\mathbf{k}} - \omega) \times \langle u_{c\mathbf{k}+\mathbf{e}_\alpha q} | u_{v\mathbf{k}} \rangle \langle u_{v\mathbf{k}} | u_{c\mathbf{k}+\mathbf{e}_\beta q} \rangle$$

where $\epsilon_{\alpha\beta}^{(2)}(\omega)$ is the dielectric function, including frequency ω, the volume of the primitive cell Ω, the Bloch vector of the incident wave **q**, *c* refers to conduction band states, *v* refers to valence band states, $u_{c\mathbf{k}}$ is the cell periodic part of the orbitals at the k-point **k**, and the vectors $\mathbf{e}_\alpha$ are unit vectors for the three Cartesian directions.[84]

ASSOCIATED CONTENT

**Supporting Information**.

Total energy of various configurations, imaginary part of the dielectric function, electronic structure, variations of bond length at polaron sites, and density of states of BiVO$_4$ as a functional of fractional α of Hartree-Fock (HF) exchange in the HSE06 functional. (PDF)

AUTHOR INFORMATION

**Corresponding Author**

**Bongjae Kim** - Department of Physics, Kyungpook National University, Daegu 41566, Korea; Email: bongjae@knu.ac.kr

**Authors**

**Seyeon Park** – Department of Physics, Kyungpook National University, Daegu 41566, Korea

**Yajing Zhang** – Department of Physics, Kyungpook National University, Daegu 41566, Korea; Max Planck Korea/POSTECH Center for Complex Phase Materials, Pohang 37673, Korea

**Michele Reticcioli** – Faculty of Physics and Center for Computational Materials Science, University of Vienna, Vienna 1090, Austria; National Research Council, CNR-SPIN, L'Aquila 67100, Italy; University of L'Aquila, L'Aquila 67100, Italy

**Cesare Franchini** – Faculty of Physics and Center for Computational Materials Science, University of Vienna, Vienna 1090, Austria; Department of Physics and Astronomy "Augusto Righi," University of Bologna, Bologna 40126, Italy

**Present Addresses**

[†]State Key Laboratory of Chemo/Biosensing and Chemometrics, Hunan University, Changsha 410082, China

**Author Contributions**

[‡]S.P. and Y.Z. contributed equally to this work.

**Notes**

The authors declare no competing financial interest.

ACKNOWLEDGMENT

This work was supported by the National Research Foundation of Korea (NRF; Grants No. NRF2021R1A4A1031920, No. RS-2021-NR061400, and No. RS2022-NR068223) and KISTI supercomputing center (project no. KSC-2023-CRE-0413). This research was funded in part by the Austrian Science Fund (FWF) 10.55776/F8100.